\documentclass{openjournal}

\usepackage[T1]{fontenc}
\usepackage{booktabs}
\setcitestyle{round}
\usepackage{xcolor}
\usepackage{hyperref}
\hypersetup{
    unicode,
    colorlinks=true,
    linkcolor=blue,
    citecolor=blue,
    urlcolor=blue
}

\begin{document}

\title{The diagnostic temperature discrepancy as evidence for non-Maxwellian coronal electrons}

\author{Victor Edmonds}
\email{vedmonds@finalstopconsulting.com}
\affiliation{Final Stop Consulting LLC}

\begin{abstract}
Two independent electron temperature diagnostics applied to the quiet solar corona yield systematically different results. Radio brightness temperatures from the Nan\c{c}ay Radioheliograph indicate $T_{\rm e} \approx 0.6$~MK, while hydrostatic scale-height modeling of the same plasma requires $T_{\rm e} \approx 1.5$~MK \citep{Mercier2015}. Both diagnostics probe electrons; they disagree by a factor of $R = 2.4 \pm 0.3$. This discrepancy persists across an eight-year dataset spanning solar minimum and is consistent with LOFAR observations at lower frequencies \citep{Vocks2018}. We consider the propagation alternative (turbulent scattering of radio emission), which operates in the correct direction to suppress the apparent brightness temperature. Comparison with the FORWARD/PSIMAS Maxwellian model \citep{Sharma2020} shows that the standard thermal structure predicts $\sim$1.6~MK; scattering accounts for the modest reduction toward observed MWA values but not the gap to 620~kK. The ratio $R$ is also invariant over the solar cycle despite measured variations in turbulence levels \citep{Gautam2024}. We propose that the residual discrepancy reflects non-Maxwellian electron velocity distributions. Radio bremsstrahlung samples the distribution core \citep{ChiuderiDrago2004, Fleishman2014}, while ionization rates and scale heights are dominated by the suprathermal tail \citep{Owocki1983}. For kappa distributions, the predicted ratio is $\kappa/(\kappa - 3/2)$; the observed $R = 2.4$ implies $\kappa \approx 2$--3. This is consistent with spectroscopic measurements in active regions \citep{Dudik2015} but in tension with perturbative theoretical predictions of $\kappa \approx 10$--25 \citep{Cranmer2014}. We make falsifiable predictions: Active Region cores should show a collapsed ratio ($R \leq 1.5$) as collisionality restores thermal equilibrium. Applying fluid transport equations (Spitzer-H\"arm conductivity) to plasmas with $\kappa \approx 2$--3 is physically invalid, but we do not compute the resulting heat flux, which remains an open problem.
\end{abstract}

\begin{keywords}
{Sun: corona, Sun: radio radiation, plasmas, methods: data analysis}
\end{keywords}

\maketitle

%==============================================================================
\section{Introduction}
\label{sec:intro}
%==============================================================================

Two electron temperature diagnostics, applied to the same quiet coronal plasma by the same instrument, disagree by a factor of 2.4. The discrepancy is stable over an eight-year solar cycle, consistent with observations from an independent observatory, and built into the full frequency-dependent spectral shape of the radio emission. We propose that this disagreement reflects the shape of the electron velocity distribution.

The argument proceeds in four steps. First, radio bremsstrahlung emissivity is dominated by electrons near the thermal velocity (the distribution core), while ionization rates and hydrostatic scale heights are dominated by suprathermal electrons (the distribution tail) \citep{Dulk1985, Owocki1983, ChiuderiDrago2004, Fleishman2014}. Second, for non-Maxwellian (kappa) distributions, the ratio of effective temperature to core temperature is $\kappa/(\kappa - 3/2)$, a standard result from kappa distribution theory. Third, the observed ratio of 2.4 implies $\kappa \approx 2$--3, overlapping with spectroscopic kappa measurements in active region loops \citep{Dudik2015, Dudik2017}, though the physical mechanisms differ: passive velocity filtration in the quiet Sun versus impulsive driving in active regions. Fourth, the framework makes falsifiable predictions: in high-density Active Region cores where collisions dominate, $\kappa \to \infty$ and the ratio should collapse toward unity. This has not been tested.

The individual physical results underlying this argument are well established. \citet{ChiuderiDrago2004} showed that kappa-distributed bremsstrahlung reduces the quiet Sun radio brightness temperature relative to Maxwellian. \citet{Owocki1983} demonstrated that ionization diagnostics applied to non-Maxwellian plasmas overestimate the electron temperature. \citet{Mercier2015} measured the factor-of-2.4 discrepancy. The present contribution is the inversion: given this specific observed ratio from the Mercier \& Chambe data, what $\kappa$ does the diagnostic discrepancy imply?

\subsection{Scope}
\label{sec:scope}

This paper identifies the diagnostic temperature discrepancy as a measurement of the electron velocity distribution shape and extracts a $\kappa$ value from published data. It demonstrates internal consistency with independent kappa measurements from EUV spectroscopy, addresses alternative explanations (turbulent scattering, ion-electron temperature separation), and presents falsifiable predictions.

This paper does not present new observations. It does not solve the coronal heating problem or compute heat transport in strongly non-Maxwellian plasmas, which remains an open problem \citep{Landi2001}. It does not claim that kappa distributions are the sole contributor to the observed discrepancy; it proposes that they provide a self-consistent framework for interpreting it.

\subsection{Plan of the paper}

Section~\ref{sec:observation} reviews the Mercier \& Chambe measurement and its spectral self-consistency. Section~\ref{sec:alternatives} addresses alternative explanations: turbulent scattering and ion-electron temperature separation. Section~\ref{sec:kappa} develops the kappa distribution framework, extracts $\kappa$ from the observed ratio, and compares with independent measurements. Section~\ref{sec:domain} establishes where this framework applies via the Knudsen number and velocity filtration. Section~\ref{sec:predictions} presents falsifiable predictions. Section~\ref{sec:heat} discusses implications for coronal heat transport. Section~\ref{sec:discussion} summarizes and identifies open questions.

%==============================================================================
\section{The observation}
\label{sec:observation}
%==============================================================================

\subsection{What Mercier \& Chambe measured}

\citet{Mercier2015} used the Nan\c{c}ay Radioheliograph (NRH) to image the quiet Sun at six frequencies between 150 and 450~MHz over 183 quiet days from 2004 to 2011, spanning nearly a full solar cycle. From the same dataset they extracted two independent products:

\begin{itemize}
\item \textbf{Brightness temperature.} Direct measurement of thermal bremsstrahlung emission from the solar disk. At frequencies below $\sim$200~MHz, where the corona is optically thick, $T_b$ saturates at approximately 620~kK.

\item \textbf{Hydrostatic scale-height temperature.} The radial density profile, measured from limb brightness falloff, is fit to a hydrostatic model. The resulting scale-height temperature $T_H$ ranges from 1350 to 1670~kK equatorially and 1650 to 2240~kK at the poles, depending on the year.
\end{itemize}

Both diagnostics probe electrons. They disagree by a factor of $R = T_H / T_b \approx 2.4 \pm 0.3$.

This is not a single-number anomaly. \citet{Mercier2015} tracked $T_H$ year by year over eight years (their Table~2). The scale-height temperature varies by roughly 20\% over the cycle, reflecting changes in coronal density structure. The brightness temperature stays locked near 620~kK. The ratio $R$ ranges from approximately 2.2 to 2.7 equatorially but never approaches unity, not even during the deep 2008--2009 solar minimum, the quietest Sun in a century.

\subsection{The spectral evidence}
\label{sec:spectral}

The discrepancy is not just a comparison of two numbers. It is built into the full frequency-dependent spectral shape.

\citet{Mercier2015} solve the radiative transfer equation using their measured density profiles (their Section~5.2). They compute what the disk brightness spectrum \emph{should} look like under two assumptions: $T_e = T_H = 1.5$~MK (the standard model) and $T_e = 620$~kK with $T_H = 1.5$~MK (a model where the electron temperature is lower than the scale-height temperature). The standard model predicts disk brightness temperatures far too high across the entire 150--445~MHz range (their Figure~8). The low-$T_e$ model fits the observed spectra (their Figures~9--12).

This spectral self-consistency is the strongest constraint. It means the discrepancy cannot be attributed to a calibration offset at a single frequency or a statistical fluctuation in a single year. Any alternative explanation must reproduce the correct frequency-dependent spectral shape across 150--445~MHz, not just a single brightness temperature.

\subsection{Independent confirmation}

\citet{Vocks2018} observed the quiet Sun with LOFAR at 25--79~MHz and found scale-height temperatures up to 2.2~MK, far exceeding brightness temperatures at those frequencies. This is the same discrepancy, measured with a different instrument at a different frequency range, during a different phase of the solar cycle. They did not discuss non-Maxwellian distributions as an explanation.

%==============================================================================
\section{Alternative explanations}
\label{sec:alternatives}
%==============================================================================

Before invoking non-Maxwellian physics, two standard alternatives must be addressed.

\subsection{Turbulent scattering}
\label{sec:scattering}

Radio waves propagating through a turbulent corona undergo scattering by electron density fluctuations. Scattering increases the apparent source size and, when the medium has non-zero optical depth, absorption along the lengthened ray paths can reduce the total emergent flux \citep{Thejappa2008}. Both effects suppress the observed brightness temperature. This is a real effect. The question is whether it accounts for the full factor-of-2.4 discrepancy.

\citet{Thejappa2008} modeled quiet Sun scattering using Monte Carlo simulations with two turbulence spectra: Kolmogorov ($\alpha = 11/3$, $\delta n/n = 10\%$) and flat ($\alpha = 3$, $\delta n/n = 2\%$). Both produce significant $T_b$ reduction. \citet{Sharma2020} used MWA observations to measure angular broadening directly, finding the quiet Sun 25--30\% larger in area than bremsstrahlung models predict, with $\delta n/n = 4.28 \pm 1.09\%$. Both the scattering cross-section and the free-free absorption coefficient scale as $\nu^{-2}$, so turbulent scattering can uniformly suppress $T_b$ across frequency without altering the spectral shape. Scattering is therefore a serious alternative that cannot be ruled out on spectral grounds alone.

The spectral degeneracy runs both ways. \citet{Dudik2012} showed that non-Maxwellian bremsstrahlung spectra in the millimeter radio range are nearly parallel to thermal spectra. Since both scattering and kappa distributions preserve the $\nu^{-2}$ spectral shape, neither can be identified from spectral slope alone. This is why the diagnostic ratio, comparing a core-sampling measurement to a tail-sampling measurement, is needed to distinguish them.

Scattering operates in the correct direction to reduce the discrepancy. Mercier \& Chambe's radiative transfer analysis (their Section~5.2), which does not include scattering, predicts that $T_e = T_H = 1.5$~MK produces disk spectra far too bright across 150--445~MHz (their Figure~8). Adding scattering on top of a 1.5~MK intrinsic model would bring the predicted brightness temperatures down toward the observed values. On spectral grounds alone, scattering is a viable explanation.

Three independent constraints limit its role.

\textbf{The FORWARD comparison.} \citet{Sharma2020} compared MWA observations with synthetic brightness temperature maps generated using the FORWARD package \citep{Gibson2016}, which computes Maxwellian thermal bremsstrahlung from the PSIMAS MHD coronal model. The FORWARD maps include the full coronal thermal structure (density, temperature, and magnetic field from data-driven MHD) but do not include scattering by density fluctuations, nor do they assume non-Maxwellian electron distributions. At 217~MHz near disk center, FORWARD predicts a peak $T_b$ of 1.61~MK; the MWA observes 1.50~MK (their Figure~4). At 162~MHz the gap widens to $\sim$25\% (1.60 vs.\ 1.20~MK), and at 108~MHz to $\sim$42\% (1.59 vs.\ 0.92~MK). The growing discrepancy at lower frequencies is consistent with scattering increasing at longer wavelengths, and the difference between FORWARD and MWA quantifies the scattering contribution.

The key point is that even at 108~MHz, where scattering produces its largest effect in the Sharma \& Oberoi dataset, the MWA peak $T_b$ is 0.92~MK --- still 50\% above the 620~kK optically thick saturation measured by \citet{Mercier2015}. Importantly, these are measurements in different optical depth regimes. The MWA observations at 108--217~MHz sample a regime where optical depth effects, thermal structure, and scattering all modulate $T_b$. Mercier \& Chambe's optically thick saturation below $\sim$200~MHz directly measures the electron temperature weighted by the emitting plasma. The standard Maxwellian model, with all thermal structure included, predicts $\sim$1.6~MK. Scattering accounts for the frequency-dependent reduction from 1.6~MK toward the MWA values. The gap between 1.6~MK and 620~kK is the discrepancy this paper interprets.

\textbf{Geometric insufficiency.} Purely flux-conserving scattering reduces $T_b$ by redistributing emission over a larger solid angle: $T_{b,\rm app} = T_{b,\rm int} \times (\Omega_{\rm int}/\Omega_{\rm app})$. To suppress $T_b$ from 1.5~MK to 0.6~MK through this mechanism alone, the apparent source area must inflate by a factor of 2.5 (i.e., $\sim$150\% broadening). \citet{Sharma2020} directly measured the quiet Sun angular broadening at 25--30\% in area. Even taking the upper end of their measured $\delta n/n$ range, the observed broadening is quantitatively insufficient by a factor of $\sim$5. As noted above, scattering with absorption can reduce $T_b$ beyond the purely geometric limit, but the FORWARD comparison places a quantitative bound on the total scattering contribution across all mechanisms.

\textbf{Solar cycle invariance.} The ratio $R$ is stable over Mercier \& Chambe's eight-year dataset, spanning both solar maximum and the deep 2008--2009 minimum (the quietest Sun in a century). Coronal turbulence levels vary with solar activity: \citet{Gautam2024} measured turbulence cascade rates from WIND spacecraft data spanning 1995--2020 and found that the cascade rate is largest during solar maximum and smallest during solar minimum. Baseline quiet-Sun turbulence persists even at solar minimum, and some coronal heating models are based on MHD turbulence. But if scattering drove the bulk of the $T_b$ suppression, $R$ should track variations in turbulence levels across the cycle. It does not. The stability of $R$, including through the deepest minimum in a century, combined with confirmation at different frequencies by LOFAR \citep{Vocks2018}, points to a cycle-invariant property of the electron distribution rather than a propagation effect.

Together, these constraints establish that scattering contributes to $T_b$ suppression but cannot account for the full discrepancy. The FORWARD comparison shows that the standard Maxwellian model with realistic thermal structure starts at $\sim$1.6~MK; scattering reduces this toward the MWA values but leaves the gap to 620~kK unexplained. The remaining factor of $\sim$2 reflects an intrinsic property of the electron distribution.

\subsection{Ion-electron temperature separation}
\label{sec:ti_te}

\citet{Mercier2015} infer $T_i \approx 2.2$~MK from the difference between $T_H$ and $T_e$, under the assumption that the electron distribution is Maxwellian. The hydrostatic scale height for a fully ionized hydrogen plasma is:
\begin{equation}
H = \frac{k_B (T_e + T_i)}{m_p g}
\label{eq:scaleheight}
\end{equation}
When observers report $T_H$, they define $T_H \equiv (T_e + T_i)/2$, so that $H = 2k_B T_H / (m_p g)$. The observed $T_H = 1.5$~MK therefore fixes $T_e + T_i = 2T_H = 3.0$~MK. In the standard Maxwellian interpretation with $T_e = T_i$, each species has $T = 1.5$~MK. When Mercier \& Chambe observe $T_b = 620$~kK and interpret it as $T_e$, they obtain $T_i = 3.0 - 0.62 = 2.38 \approx 2.2$~MK. Could this ion-electron temperature separation, rather than a non-Maxwellian electron distribution, explain the observed ratio?

Two problems arise. First, the inference is circular. The value $T_i \approx 2.2$~MK is derived under the Maxwellian assumption. If the electron distribution is non-Maxwellian, the pressure (and therefore the scale height) depends on $T_{\rm eff}$ rather than $T_e$. The observational constraint becomes $T_{\rm eff} + T_i = 2T_H = 3.0$~MK, and the kappa value follows from $\kappa/(\kappa - 3/2) = T_{\rm eff}/T_{\rm core}$. Two limiting cases bracket the range:

\begin{itemize}
\item \textbf{Core equipartition} ($T_i = T_{\rm core} = 0.6$~MK): The constraint gives $T_{\rm eff} = 3.0 - 0.6 = 2.4$~MK, so $T_{\rm eff}/T_{\rm core} = 4.0$, yielding $\kappa = 2.0$.
\item \textbf{Effective equipartition} ($T_i = T_{\rm eff}$): The constraint gives $2T_{\rm eff} = 3.0$~MK, so $T_{\rm eff} = 1.5$~MK, $T_{\rm eff}/T_{\rm core} = 2.5$, and $\kappa = 2.5$.
\end{itemize}

The kappa inference is robust across the full range of $T_i$ assumptions: $\kappa = 2.0$--2.5. The core equipartition case is particularly noteworthy: $\kappa = 2.0$ sits exactly on the mathematical boundary where the third velocity moment (heat flux) diverges (Section~\ref{sec:heat}).

Second, the solar cycle test fails. \citet{Landi2007} showed that the ion-electron temperature difference in the corona varies strongly with solar activity: $T_i - T_e$ is substantial near solar maximum but ``nearly absent'' at the 1996 minimum. If $T_i \gg T_e$ drove the diagnostic ratio $R$, then $R$ should approach unity at solar minimum. Mercier \& Chambe's data through the 2008--2009 minimum show $R$ unchanged.

The standard interpretation of the discrepancy requires two unexplained phenomena: $T_e \ll T_H$ and $T_i \gg T_e$. The kappa interpretation explains the diagnostic disagreement with one mechanism. $T_i \gg T_e$ and non-Maxwellian electrons are not mutually exclusive; both may contribute. But the cycle invariance of $R$, despite documented variation in $T_i/T_e$, indicates that kappa provides a baseline contribution that ion-electron separation alone cannot explain.

%==============================================================================
\section{Kappa distributions and temperature amplification}
\label{sec:kappa}
%==============================================================================

\subsection{The kappa distribution}

Non-equilibrium plasmas are characterized by kappa distributions \citep{Livadiotis2009, Pierrard2010}, which exhibit power-law tails rather than the exponential cutoff of a Maxwellian:
\begin{equation}
f_\kappa(v) \propto \left(1 + \frac{v^2}{\kappa \theta^2}\right)^{-(\kappa+1)}
\label{eq:kappa}
\end{equation}
where $\theta^2 = 2k_B T_{\rm core}/m$ and $\kappa$ parameterizes the departure from equilibrium.\footnote{We use the ``standard'' kappa formulation in which $T_{\rm core}$ is the most-probable-speed temperature and depends on $\kappa$ through Equation~\ref{eq:amplification}. An alternative ``regularized'' formulation \citep{Livadiotis2009} defines the temperature parameter such that $\langle E \rangle$ depends only on $T$ and not on $\kappa$. The two conventions describe the same distributions with different labeling. In the regularized convention, $T = T_{\rm eff}$ throughout, and the distinction between core and effective temperatures is absorbed into the definition of $\theta$. All physical results in this paper are independent of the convention chosen.} As $\kappa \to \infty$, the Maxwellian is recovered. At low $\kappa$, the distribution has an enhanced high-energy tail: more electrons at speeds well above the thermal velocity, fewer near the peak.

\begin{figure}[t]
\centering
\includegraphics[width=\textwidth]{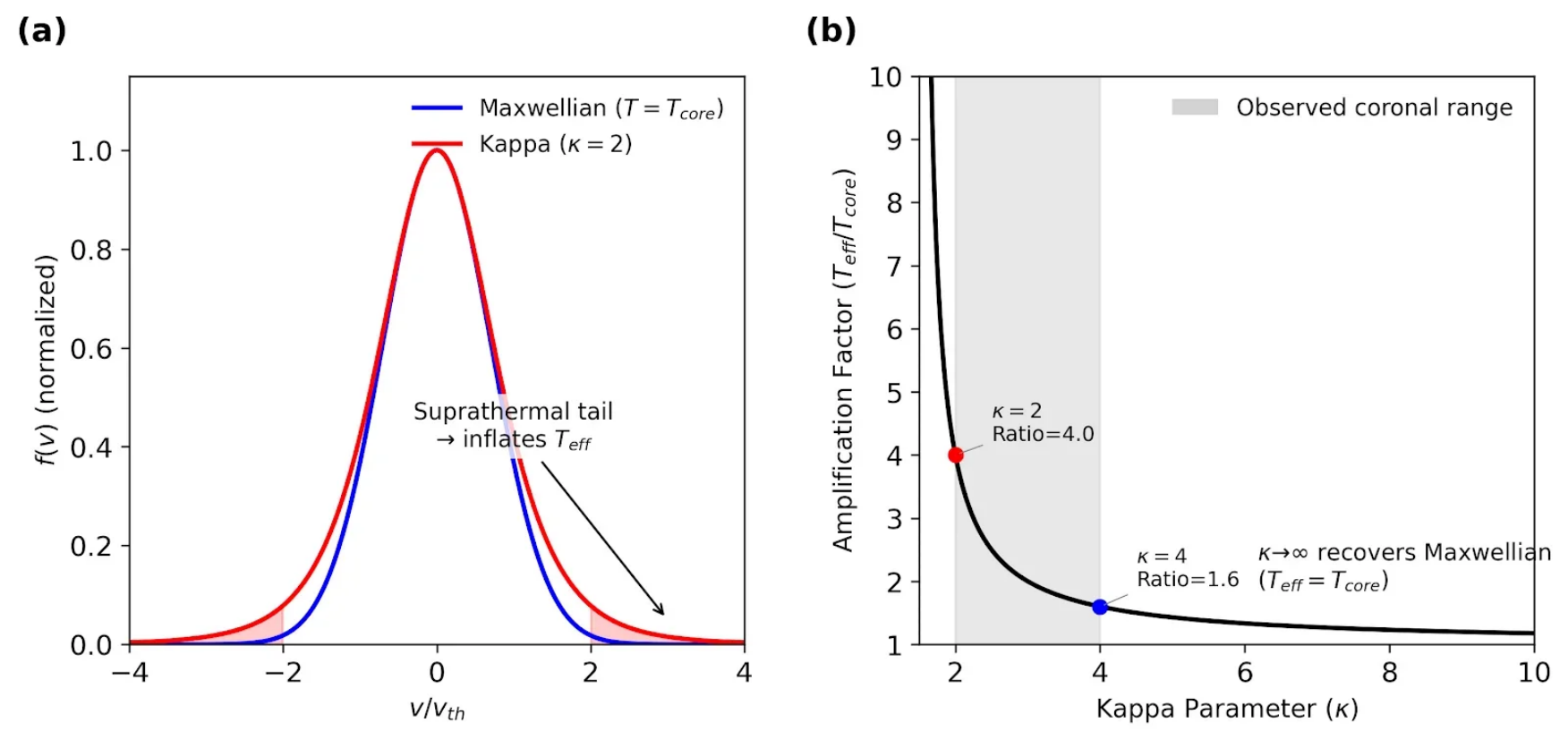}
\caption{(\textit{a})~Normalized velocity distributions for a Maxwellian (blue) and a kappa distribution with $\kappa = 2$ (red), both at the same core temperature $T_{\rm core}$. The enhanced suprathermal tail (shaded) inflates the effective temperature $T_{\rm eff}$ measured by ionization and scale-height diagnostics, while radio bremsstrahlung samples the thermal core. (\textit{b})~Temperature amplification factor $T_{\rm eff}/T_{\rm core} = \kappa/(\kappa - 3/2)$ as a function of $\kappa$. The grey band marks the observed coronal range $\kappa \approx 2$--3 inferred from the Mercier \& Chambe (2015) diagnostic ratio $R = 2.4$.}
\label{fig:kappa}
\end{figure}

The effective temperature, defined via the second moment of the distribution, is:
\begin{equation}
T_{\rm eff} = T_{\rm core} \cdot \frac{\kappa}{\kappa - 3/2}
\label{eq:amplification}
\end{equation}
This is the temperature that governs processes sensitive to the mean kinetic energy (pressure, scale height) or to the suprathermal population (ionization, recombination). The core temperature $T_{\rm core}$ governs processes dominated by the thermal peak. Figure~\ref{fig:kappa} illustrates both the distribution shape and the amplification factor as a function of $\kappa$.

\subsection{How different diagnostics weight the distribution}
\label{sec:weighting}

Different measurements sample different parts of the same velocity distribution.

\textbf{Radio bremsstrahlung.} Thermal bremsstrahlung emissivity at radio frequencies is dominated by electrons near the thermal velocity. \citet{ChiuderiDrago2004} computed bremsstrahlung emissivity and absorption coefficients for kappa distributions in the quiet Sun and showed that $T_b$ is lower for all $\kappa$ values than for a pure Maxwellian at the same effective temperature. \citet{Fleishman2014} derived comprehensive free-free emissivity and absorption coefficients for kappa distributions, from which it follows that the brightness temperature of optically thick kappa bremsstrahlung approximates $T_{\rm core}$, not $T_{\rm eff}$.

\textbf{Ionization rates.} Collisional ionization has threshold energies typically much larger than $k_B T$ and cross-sections that increase with electron energy above threshold. \citet{Owocki1983} demonstrated that ionization rates in non-Maxwellian plasmas are dominated by suprathermal electrons. Fitting a Maxwellian ionization model to kappa-distributed plasma yields an apparent temperature $\approx T_{\rm eff}$. (Owocki \& Scudder used a general parameterized non-Maxwellian distribution rather than kappa specifically, but the result generalizes to kappa and is universally cited in the kappa literature.)

\textbf{Hydrostatic scale height.} The pressure scale height $H = k_B T / (m_p g)$ reflects the mean kinetic energy of the plasma. For a kappa distribution, this corresponds to $T_{\rm eff}$ rather than $T_{\rm core}$.

The predicted diagnostic ratio is therefore:
\begin{equation}
R \equiv \frac{T_{\rm scale\text{-}height}}{T_{\rm brightness}} \approx \frac{T_{\rm eff}}{T_{\rm core}} = \frac{\kappa}{\kappa - 3/2}
\label{eq:ratio}
\end{equation}

\subsection{Optical depth self-consistency}
\label{sec:tau}

An immediate concern is whether the kappa model preserves optical thickness. If the corona became optically thin at the lower core temperature, the brightness temperature would no longer equal the electron temperature, invalidating the entire framework.

The Maxwellian free-free absorption coefficient scales as $\alpha_\nu \propto n_e^2 T^{-3/2} \nu^{-2}$. Using this as a bounding estimate, replacing $T_{\rm eff} = 1.5$~MK with $T_{\rm core} = 0.6$~MK at fixed density and frequency, changes the optical depth by:
\begin{equation}
\frac{\tau_\kappa}{\tau_{\rm standard}} \gtrsim \left(\frac{T_{\rm eff}}{T_{\rm core}}\right)^{3/2} = (2.5)^{1.5} \approx 4
\end{equation}
We use the Maxwellian scaling as a lower bound; \citet{Fleishman2014} compute the exact $\kappa$-dependent absorption coefficient, which includes additional enhancement from the non-Maxwellian tail at low $\kappa$. The kappa model is therefore \emph{more} optically thick than the standard model. If the standard model is marginally opaque ($\tau \approx 1$), the kappa model is solidly opaque ($\tau \gtrsim 4$). The radio brightness temperature measurement remains valid.

We note that the relationship $T_b \approx T_{\rm core}$ is a leading-order approximation. The exact brightness temperature of optically thick kappa bremsstrahlung depends on the frequency-dependent emissivity-to-absorption ratio, which \citet{Fleishman2014} show departs slightly from $T_{\rm core}$ at low $\kappa$. For $\kappa \approx 2.5$, the correction is of order 10--20\%, small compared to the factor-of-2.4 discrepancy under consideration.

\subsection{Inferring kappa from the observed ratio}

Inverting Equation~\ref{eq:ratio}:
\begin{equation}
\kappa = \frac{3R}{2(R-1)}
\label{eq:inversion}
\end{equation}

For the observed $R = 2.4 \pm 0.3$, this yields $\kappa = 2.57^{+0.29}_{-0.19}$.\footnote{Error propagation: $|d\kappa/dR| = 3/[2(R-1)^2]$. At $R = 2.4$, $|d\kappa/dR| = 0.765$ and $\sigma_\kappa = 0.765 \times 0.3 = 0.23$. Since the inversion is nonlinear, we also compute the exact bounds: $R = 2.1 \Rightarrow \kappa = 2.86$; $R = 2.7 \Rightarrow \kappa = 2.38$.} The range $\kappa \approx 2.4$--2.9 is robust.

\subsection{Comparison with independent measurements}
\label{sec:comparison}

Table~\ref{tab:kappa} collects $\kappa$ measurements from different methods and environments.

\begin{table}[ht]
\centering
\caption{Kappa values from different diagnostics and environments.}
\label{tab:kappa}
\begin{tabular}{lllcl}
\toprule
Method & Environment & $\kappa$ & Mechanism & Reference \\
\midrule
$T_b/T_H$ ratio & Quiet Sun & 2--3 & Passive (filtration) & This work \\
Fe line ratios & AR loops & 2--5 & Driven (impulsive) & Dud\'{\i}k+ 2015, 2017 \\
In situ (halo) & Solar wind, 1~AU & 2.5--7.8 & Evolving & \v{S}tver\'ak+ 2009 \\
In situ & Near Sun, 0.17~AU & $\sim$5--8 & Evolving & Halekas+ 2020 \\
Perturbative theory & Low corona & 10--25 & Wave-driven & Cranmer 2014 \\
\bottomrule
\end{tabular}
\end{table}

Several points merit discussion.

\textbf{Active region EUV measurements.} The overlap between our quiet-Sun $\kappa \approx 2$--3 and the active-region values of $\kappa \approx 2$--5 from \citet{Dudik2015, Dudik2017} may reflect distinct physical origins for similar $\kappa$ values. In the quiet Sun, low densities permit velocity filtration \citep{Scudder1992a, Scudder1992b} to generate persistent suprathermal tails, a passive departure from equilibrium maintained by the temperature gradient. In active region loops, impulsive energy release from reconnection or nanoflares continuously regenerates tails faster than collisions can thermalize them, a driven departure. Both mechanisms can produce similar $\kappa$ values through different physics.

\textbf{Solar wind in situ measurements.} \citet{Stverak2009} fit a three-component model (bi-Maxwellian core + bi-kappa halo + strahl) to Helios, Cluster, and Ulysses data. The halo component shows $\kappa$ \emph{decreasing} from $\sim$7.8 at 0.4~AU to $\sim$2.5 at 4~AU, becoming more non-Maxwellian with distance as strahl electrons scatter into the halo during transit. The extrapolation toward the corona is ambiguous: at the closest approach (0.3~AU), halo $\kappa \approx 7$--8, which does not directly constrain the coronal value. Parker Solar Probe measurements closer to the Sun \citep{Halekas2020} suggest $\kappa \sim 5$--8 at 0.17~AU, broadly consistent with a more Maxwellian near-Sun distribution that evolves outward.

\textbf{Theoretical predictions.} \citet{Cranmer2014} modeled nonlocal electron transport in wave/turbulence-driven coronal models using perturbative methods, predicting $\kappa \approx 10$--25 in the low corona. At $\kappa = 15$, the amplification factor $T_{\rm eff}/T_{\rm core}$ is only 1.11, far short of the observed 2.4. This is a genuine theory-observation tension. The perturbative approach may underestimate the degree of non-thermality if velocity filtration (fundamentally non-perturbative, driven by the exponential sensitivity of particle transmission through the transition region potential) dominates over wave-driven effects in the quiet Sun. Resolving this tension requires either non-perturbative theoretical calculations or independent empirical constraints on coronal $\kappa$.

\subsection{The EUV detection paradox}
\label{sec:euv}

If the quiet Sun has $\kappa \approx 2$--3, why don't EUV line ratio diagnostics detect it? More pointedly: five independent EUV-based methods (emission measure loci, line ratios, DEM inversions, density-sensitive ratios, and EUV-to-UV spanning ratios) all yield temperatures near 1.5~MK \citep{DelZanna2018}. If the core electron temperature is really 0.6~MK, why does every EUV method agree on 1.5~MK?

All five methods are downstream of a single physical bottleneck: collisional ionization.

\textbf{The ionization gatekeeper.} The charge state distribution of coronal iron is set by collisional ionization and radiative recombination. The ionization potentials of the dominant quiet Sun ions (Fe~IX through Fe~XII) range from 234 to 331~eV. At a core temperature of $T_{\rm core} \approx 0.6$~MK, the thermal energy is $k_B T_{\rm core} \approx 52$~eV. These ionization thresholds are 4.5--6.4 times the thermal energy. In a Maxwellian, the ionizing electrons are drawn from the far exponential tail. In a kappa distribution with $\kappa \approx 2$--3, the enhanced power-law tail provides substantially more ionizing electrons than a Maxwellian at $T_{\rm core}$, but a comparable number to a Maxwellian at $T_{\rm eff} \approx 1.5$~MK. \citet{Owocki1983} established this general result for non-Maxwellian distributions: ionization rates are dominated by suprathermal electrons and therefore track $T_{\rm eff}$, not $T_{\rm core}$.

The charge state distribution is therefore set at $T_{\rm eff}$. This is not an artifact of assuming Maxwellian; it is a direct physical consequence of the tail domination of threshold-crossing processes.

\textbf{Line ratio degeneracy.} Once the charge states are determined, the EUV line ratios within each ion depend on excitation rates, which are far less sensitive to the distribution shape. \citet{Dudik2014} computed synthetic spectra for kappa distributions and showed that the vast majority of strong coronal EUV line ratios vary by less than 20\% across $\kappa = 2$--25. The reason is straightforward: excitation energies for allowed transitions within a given ion are comparable to the thermal energy (a few times $k_B T$), so the excitation rates sample a broad range of electron speeds near the thermal peak, not the far tail. Intra-ion line ratios are therefore nearly degenerate in $\kappa$.

\textbf{Cross-diagnostic blindspot.} EM loci, DEM inversions, and density-sensitive ratios all interpret emission from ions whose populations are already determined by $T_{\rm eff}$. The line intensities are proportional to the ion fraction (set by $T_{\rm eff}$) multiplied by excitation rates (nearly $\kappa$-independent). Every EUV method therefore converges on $T_{\rm eff} \approx 1.5$~MK, regardless of $\kappa$. The agreement among the five methods is not independent confirmation of a Maxwellian; it is a consequence of their shared dependence on ionization, which is itself tail-dominated.

\textbf{Radio breaks the degeneracy.} Bremsstrahlung (free-free emission) involves no ionization thresholds. The emissivity is an integral over the full electron velocity distribution weighted by the Coulomb cross-section, which has no energy threshold and peaks near the thermal velocity. Radio brightness temperature therefore measures $T_{\rm core}$, not $T_{\rm eff}$. It is the one coronal temperature diagnostic that bypasses the ionization gatekeeper entirely.

This resolves the apparent paradox. Under a kappa distribution with $\kappa \approx 2$--3: all EUV methods yield $\approx T_{\rm eff} \approx 1.5$~MK (because ionization sets the charge states at $T_{\rm eff}$), radio yields $\approx T_{\rm core} \approx 0.6$~MK (because bremsstrahlung samples the distribution core), and the two disagree by $R = \kappa/(\kappa - 3/2)$. The EUV convergence at 1.5~MK is not evidence against kappa distributions; it is a prediction of the kappa framework.

We note two caveats. First, \citet{Dudik2014} identified specific exceptions to the general insensitivity: the Fe~IX forbidden line at 6378~\AA\ increases by up to a factor of two at low $\kappa$. Systematic quiet Sun observations of this line would provide an independent test. Second, a full quantitative verification (forward-modeling the complete EUV spectrum under kappa distributions using CHIANTI atomic data and comparing with observed quiet Sun intensities) has not been performed. We leave this calculation to researchers with the spectroscopic expertise to undertake it. The argument presented here is structural: it shows why EUV convergence at $T_{\rm eff}$ is expected under kappa.

%==============================================================================
\section{Domain of applicability}
\label{sec:domain}
%==============================================================================

\subsection{The Knudsen number}

Non-Maxwellian distributions persist only where collisions are too infrequent to enforce thermal equilibrium. The relevant dimensionless parameter is the Knudsen number:
\begin{equation}
\mathrm{Kn} = \frac{\lambda_{\rm mfp}}{L_T}
\label{eq:knudsen}
\end{equation}
where $\lambda_{\rm mfp}$ is the electron mean free path and $L_T$ is the temperature scale height. When $\mathrm{Kn} \gtrsim 0.01$, the classical diffusive (Spitzer-H\"arm) approximation breaks down and kinetic effects dominate \citep{Landi2001, Dorelli2003, Boldyrev2019}.

For the quiet corona ($n_e \sim 10^8$~cm$^{-3}$, $T \sim 10^6$~K): $\lambda_{\rm mfp} \sim 10^7$~cm, $L_T \sim 10^{8}$--$10^{9}$~cm, giving $\mathrm{Kn} \sim 0.01$--0.1. This is the kinetic regime. The critical point, noted by \citet{Shoub1983}, is that the mean free path scales as $v^4$: suprathermal electrons at a few times the thermal velocity have mean free paths orders of magnitude longer than bulk electrons. Even in regions where the bulk is marginally collisional, the tail population is effectively collisionless.

For Active Region cores ($n_e \sim 10^{9}$--$10^{10}$~cm$^{-3}$), $\mathrm{Kn} \lesssim 0.01$, approaching the fluid regime where Maxwellian distributions are expected.

\subsection{Velocity filtration: the generation mechanism}

The physical mechanism generating kappa distributions in the quiet corona is velocity filtration, identified by \citet{Scudder1992a, Scudder1992b}.

The transition region has steep temperature gradients ($L_T \sim 10^8$~cm, $\mathrm{Kn} \sim 0.1$). In this weakly collisional boundary, electrons are not in thermal equilibrium with the local gas. Fast electrons from the chromosphere preferentially pass through the gravitational and electric potential barriers at the transition region because their kinetic energy exceeds the potential energy, the same mechanism that generates the solar wind electron strahl. The resulting coronal distribution has an enhanced suprathermal tail relative to Maxwellian at the same mean energy.

This is a passive departure from equilibrium. No impulsive energy input is required beyond what maintains the temperature gradient. The distribution propagates into the extended corona, where low collision frequencies ($\nu_{ei} \sim 7$~s$^{-1}$ for thermal electrons) preserve the non-thermal tail over the thermal timescale. \citet{Scudder2019} showed that the degree of non-thermality ($\kappa$) is directly linked to the dimensionless parallel electric field, itself a measure of departure from local thermodynamic equilibrium. The filtration mechanism is not specific to the Sun: \citet{MeyerVernet1995} identified its signature in the Io plasma torus, and \citet{Pierrard1996} in the terrestrial ionosphere. It operates wherever a plasma traverses a potential gradient with a finite Knudsen number.

Recent kinetic models confirm this picture. \citet{Vocks2008} demonstrated self-consistent formation of suprathermal distributions in quiet coronal loops via whistler wave interaction. \citet{Barbieri2024} showed that intermittent chromospheric heating events generate coronal electron velocity distributions with shrinking thermal cores and dominant suprathermal tails: velocity filtration from first principles, without imposing kappa distributions a priori. \citet{Barbieri2025} extended this to three-dimensional kinetic models with spatially intermittent heating.

%==============================================================================
\section{Predictions}
\label{sec:predictions}
%==============================================================================

\subsection{Active Region core collapse}
\label{sec:collapse}

In the quiet Sun, the mechanism maintaining low $\kappa$ is velocity filtration, a steady-state process sustained by the Knudsen number. No impulsive energy input is assumed. The suprathermal tail is generated at the transition region boundary and preserved by low collisionality in the extended corona.

In Active Region cores ($n_e \geq 10^9$~cm$^{-3}$), collisionality increases. The average collision frequency scales as $\nu \propto n_e T^{-3/2}$, but for individual electrons the rate scales as $v^{-3}$ \citep{Shoub1983}: low-velocity (thermal core) electrons thermalize far faster than high-velocity (suprathermal tail) electrons. The bulk of the distribution therefore relaxes toward Maxwellian while residual suprathermal tails persist on longer timescales. The resulting distribution is no longer a clean kappa --- it is closer to a Maxwellian core plus a residual power-law tail, analogous to the solar wind core+halo decomposition.

In this partially thermalized regime, each diagnostic responds differently. Radio bremsstrahlung still samples the bulk, but the bulk has thermalized upward --- $T_{\rm core}$ has relaxed toward $T_{\rm eff}$ --- so $T_b$ increases relative to the quiet Sun case. The scale height and ionization balance remain sensitive to the tail (if it persists), so $T_H$ stays approximately constant. $R$ therefore decreases because $T_b$ has increased while $T_H$ has not, but $R$ does not reach unity as long as a residual tail keeps $T_H$ above $T_b$.

\textbf{Prediction:} $R$ should decrease systematically with increasing density, tracing the transition from tail-dominated to collision-dominated regimes. Different AR structures at different densities --- quiet Sun, AR periphery, AR core --- should trace out a continuous $R(n_e)$ curve that maps the competition between tail regeneration and collisional thermalization. In AR cores with quasi-steady heating \citep{Antiochos2003}, $R$ should fall well below the quiet Sun value of 2.4.

\textbf{Falsification:} If the diagnostic ratio in AR cores remains $\geq 2.2$ (indistinguishable from quiet Sun), this framework fails, because even partial thermalization should produce measurable collapse.

The kappa inversion formula $\kappa = 3R/[2(R-1)]$ assumes a kappa distribution. In the partially thermalized regime, the distribution is a hybrid. The formula still yields a number, but that number is an effective $\kappa$ characterizing the diagnostic ratio rather than the distribution shape. It remains useful as a single-parameter description of how non-Maxwellian the plasma appears to the core-versus-tail comparison, but it does not map cleanly onto the distribution function in this intermediate regime.

This prediction applies specifically to AR cores with quasi-steady heating --- ``quasi-steady'' referring to the time-averaged heating rate being approximately constant \citep{Antiochos2003}, not the absence of dynamic activity. Small-scale reconnection events are ubiquitous, but the time-averaged collisionality they produce determines whether the tail regeneration rate can sustain a non-Maxwellian distribution. The prediction does not apply to dynamic AR loops or transient features, where impulsive energy release can maintain low $\kappa$ despite high densities. In AR cores, gyroresonance emission at low harmonics of the electron cyclotron frequency can dominate over bremsstrahlung, complicating the interpretation of radio brightness temperatures. The test therefore requires either frequencies above the gyroresonance layers or careful isolation of the free-free component.

The distinction between quiet Sun, AR loops, and AR cores is central. Table~\ref{tab:regimes} summarizes the three regimes.

\begin{table}[ht]
\centering
\caption{Three regimes of coronal electron distribution shape.}
\label{tab:regimes}
\begin{tabular}{lllcl}
\toprule
Environment & Mechanism & Driving & Collisionality & Expected $\kappa$ \\
\midrule
Quiet Sun & Velocity filtration & Passive & Low (Kn $> 0.01$) & 2--3 \\
AR loops & Reconnection/nanoflares & Impulsive & Moderate & 2--5 (maintained) \\
AR cores & Quasi-steady heating & None & High (Kn $< 0.01$) & $\gg 5$ (Maxwellian) \\
\bottomrule
\end{tabular}
\end{table}

Supporting evidence for the collisionality dependence exists from flare observations. \citet{Dzifcakova2018} showed $\kappa$ evolving from $\lesssim 2$ during the impulsive phase toward progressively larger values during the gradual phase. Once impulsive driving ceases, collisions control the distribution shape.

The apparent paradox noted by \citet{Lorincik2020}, low $\kappa$ in AR loops despite elevated density, is precisely this distinction. AR loops are dynamic structures where impulsive driving dominates over collisional thermalization. AR cores, with their quasi-steady heating and high density, are the environment where thermalization should win. The prediction targets AR cores, not AR loops.

\subsection{Topological control}

Open magnetic field lines allow suprathermal electrons to escape before thermalizing, depleting the high-energy tail. For tail electrons at $v \sim 10\,v_{\rm th}$, the collision time ($\tau_{\rm coll} \propto v^3$) exceeds the escape time along open field lines. Closed-field regions, by contrast, trap suprathermal electrons, preserving the enhanced tail.

\textbf{Prediction:} At matched electron density, closed-field regions should show a larger diagnostic discrepancy than open-field regions.

This is consistent with Parker Solar Probe observations of the electron strahl \citep{Halekas2020}, which provide direct evidence of suprathermal escape along open field lines.

%==============================================================================
\section{Implications for coronal heat transport}
\label{sec:heat}
%==============================================================================

This paper does not attempt to solve the coronal heating problem. If the $R = 2.4$ discrepancy is real and reflects $\kappa \approx 2$--3, then calculating conductive losses via classical fluid approximations is physically invalid.

Standard coronal models use Spitzer-H\"arm conductivity, $q_{\rm SH} = \kappa_0 T^{7/2}/L_T$, with $T = T_H \approx 1.5$~MK as input. This assumes two things: a Maxwellian electron velocity distribution and a collision-dominated regime ($\mathrm{Kn} \ll 0.01$). In the quiet corona, neither assumption holds. The Knudsen number is $\sim$0.01--0.1 (Section~\ref{sec:domain}), and the distribution is far from Maxwellian if $\kappa \approx 2$--3.

An objection arises immediately. One might expect that kappa distributions, which provide more suprathermal electrons, would \emph{increase} heat flux relative to Maxwellian, since suprathermals are the primary heat carriers. This expectation is correct at fixed temperature. But it misses the deeper problem: the dominant error in the standard calculation is using the wrong input temperature. If $T_{\rm eff} = 1.5$~MK but $T_{\rm core} = 0.6$~MK, plugging 1.5~MK into the Spitzer-H\"arm formula uses a temperature that doesn't correspond to the bulk of the distribution. The $T^{7/2}$ dependence means the sensitivity to this error is extreme.

Moreover, \citet{Landi2001} showed through kinetic simulations that at $\kappa < 5$, heat flux can reverse direction, flowing from cold to hot regions. At $\kappa \approx 2$--3, the heat flux problem is not merely quantitative (``Spitzer overestimates by factor X'') but qualitative: the sign and even the direction of energy flow depend on the distribution shape in ways that fluid closures cannot capture.

A further constraint sharpens the picture. For $\kappa \leq 2$, the third velocity moment of the kappa distribution (the moment that defines the heat flux in kinetic theory) diverges mathematically. Fluid transport models are not merely inaccurate in this regime; they are undefined. Our inferred $\kappa \approx 2.4$--2.9 sits just above this divergence threshold, in a regime where the heat flux integral converges but is dominated by the high-energy tail rather than the bulk. Standard fluid closures, which truncate the moment hierarchy assuming near-Maxwellian distributions, cannot capture this physics.

We do not compute the actual conductive flux for $\kappa \approx 2$--3 plasmas. The Landi \& Pantellini result applies to specific boundary conditions that may not directly map to the coronal geometry. The quantitative impact on the coronal energy budget remains unresolved. What we can say is that applying any macroscopic fluid transport equation to plasma in this regime is not physically justified.

The implications are limited but specific: conductive losses calculated from fluid models with $T_{\rm eff} = 1.5$~MK as input are suspect. The portion of the coronal heating problem attributable to conduction may be different (larger or smaller) than currently assumed. This is an open problem, not a solved one.

%==============================================================================
\section{Discussion and conclusions}
\label{sec:discussion}
%==============================================================================

\subsection{Summary}

The diagnostic temperature discrepancy in the quiet solar corona ($R = T_H/T_b \approx 2.4$) is not a calibration artifact. It is built into the frequency-dependent spectral shape of the radio emission \citep{Mercier2015}, stable over an eight-year solar cycle, and consistent with LOFAR observations at lower frequencies \citep{Vocks2018}. Turbulent scattering contributes to $T_b$ suppression but cannot account for the full discrepancy: the standard Maxwellian model with realistic thermal structure predicts $\sim$1.6~MK \citep{Sharma2020, Gibson2016}; scattering reduces this toward observed MWA values but leaves the gap to 620~kK unexplained (Section~\ref{sec:scattering}). Ion-electron temperature separation fails the solar cycle test (Section~\ref{sec:ti_te}).

We identify the discrepancy as a measurement of the electron velocity distribution shape. Radio bremsstrahlung samples the distribution core; ionization and scale-height diagnostics sample the suprathermal tail. For kappa distributions, the predicted ratio $\kappa/(\kappa - 3/2)$ matches the observed $R = 2.4$ at $\kappa \approx 2$--3.

\subsection{Relation to previous work}

\citet{ChiuderiDrago2004} established the theoretical framework: kappa-distributed bremsstrahlung reduces quiet Sun radio brightness temperature relative to Maxwellian. Their work predates the Mercier \& Chambe measurement and focused on microwave/cm-mm wavelengths without framing the result as a diagnostic ratio or extracting a specific $\kappa$. Our contribution is the inversion: applying their physics to Mercier \& Chambe's specific data to extract $\kappa$ from the observed discrepancy.

\citet{Fleishman2014} provide the comprehensive kappa emissivity and absorption framework that underpins the bremsstrahlung side of the argument. \citet{Owocki1983} established the ionization side.

More broadly, the empirical $\kappa \approx 2$--3 inferred here from diagnostic ratios converges on the prediction of \citet{Scudder1992a, Scudder1992b} from the opposite direction. Scudder argued on theoretical grounds that velocity filtration through the transition region potential must generate non-Maxwellian coronal distributions; the present work provides a quantitative measurement of the degree of that departure, three decades later, using observational data that were not available in 1992.

The tension with \citet{Cranmer2014}, who predicts $\kappa \approx 10$--25 from perturbative transport theory, is real and unresolved. Theoretical models predict mild suprathermal tails; our empirical result implies strong ones. Resolving this likely requires non-perturbative calculations of velocity filtration in realistic coronal geometries, a direction that \citet{Scudder2019} has begun to develop through the Steady Electron Runaway Model framework.

\subsection{Open questions}

We identify five questions that this paper raises but does not answer:

\begin{enumerate}
\item \textbf{Heat transport.} What is the actual heat flux in a closed-field coronal structure with $\kappa \approx 2$--3? \citet{Landi2001} show flux reversal at low $\kappa$, but the coronal boundary conditions differ from their simulation geometry.

\item \textbf{Theory-observation tension.} Why does perturbative transport theory predict $\kappa \approx 10$--25 \citep{Cranmer2014} while the diagnostic ratio implies $\kappa \approx 2$--3? Is velocity filtration the missing non-perturbative element?

\item \textbf{EUV confirmation.} Can $\kappa$-sensitive forbidden-line diagnostics \citep{Dudik2014} independently confirm low $\kappa$ in the quiet Sun, where density-diagnostic EUV ratios are insensitive?

\item \textbf{Scattering decomposition.} The FORWARD comparison (Section~\ref{sec:scattering}) provides a first-order decomposition of scattering and intrinsic effects at 108--217~MHz. Extending this into the optically thick regime below $\sim$200~MHz, where the diagnostic discrepancy is measured, requires forward-modeling with scattering included in the radiative transfer. Does the residual discrepancy persist when scattering is self-consistently incorporated?

\item \textbf{Emission measure reconciliation.} If the bulk electron temperature is $T_{\rm core} \approx 0.6$~MK, collisional excitation rates for EUV lines are lower than at 1.5~MK in a Maxwellian. Matching the observed absolute EUV radiances would therefore require a higher coronal emission measure ($\mathrm{EM} = \int n_e^2\, dl$) than inferred under Maxwellian assumptions. We note that the enhanced suprathermal tail of a $\kappa \approx 2$--3 distribution partially compensates for the cold core: excitation rate coefficients evaluated at $T_{\rm eff}$ are expected to be comparable to Maxwellian rates at the same temperature \citep{Dudik2014}, mitigating the required EM adjustment. Nevertheless, a quantitative comparison of absolute line radiances under kappa distributions with white-light electron densities and volumetric filling factors has not been performed.
\end{enumerate}

\subsection{Observational tests}

The framework makes specific, testable predictions:

\begin{itemize}
\item \textbf{AR core collapse.} Radio brightness measurements of AR cores at meter wavelengths (NRH, LOFAR, MWA) should show $R \leq 1.5$, in contrast to the quiet Sun $R \approx 2.4$.

\item \textbf{Density dependence.} Simultaneous multi-diagnostic campaigns (radio brightness + spectroscopic density profiles) should reveal $R$ decreasing systematically with $n_e$.

\item \textbf{Topological control.} At matched electron density, closed-field regions should show larger $R$ than open-field regions.

\item \textbf{Forbidden-line detection.} Systematic quiet-Sun observations of the Fe~IX 6378~\AA\ forbidden line should reveal intensity enhancements consistent with low $\kappa$.
\end{itemize}

\subsection{Conclusions}

\begin{enumerate}
\item The factor-of-2.4 diagnostic temperature discrepancy in the quiet corona is spectrally self-consistent, stable over a solar cycle, and confirmed by independent observations. It is not a calibration artifact.

\item Interpreted as a kappa distribution measurement, the discrepancy implies $\kappa \approx 2$--3 for quiet Sun electrons. This is consistent with spectroscopic measurements in active regions but in tension with perturbative theoretical predictions.

\item The framework predicts that the diagnostic ratio should collapse toward unity in high-density Active Region cores. This prediction is falsifiable with existing instruments.

\item Applying Spitzer-H\"arm conductivity to plasma with $\kappa \approx 2$--3 is physically invalid. The quantitative impact on the coronal energy budget is an open problem.

\item The diagnostic discrepancy is a measurement of the electron velocity distribution shape, not a problem to be solved.
\end{enumerate}

\section*{Acknowledgements}

This work made use of published observations from the Nan\c{c}ay Radioheliograph and LOFAR solar imaging programs. The author thanks the solar physics community for maintaining open access to data and fostering accessible scientific discourse.

\section*{Data availability}
This work interprets previously published observational results. No new data were generated or analysed. The primary observational constraints are from \citet{Mercier2015} and \citet{Vocks2018}.

\bibliographystyle{plainnat}

\begin{thebibliography}{99}

\bibitem[Antiochos et~al.(2003)]{Antiochos2003}
Antiochos, S.~K., Karpen, J.~T., DeLuca, E.~E., Golub, L., \& Hamilton, P. (2003).
\newblock Constraints on Active Region Coronal Heating.
\newblock \textit{The Astrophysical Journal}, 590, 547.

\bibitem[Barbieri et~al.(2024)]{Barbieri2024}
Barbieri, L., Casetti, L., Verdini, A., \& Landi, S. (2024).
\newblock Temperature inversion in a gravitationally bound plasma: Case of the solar corona.
\newblock \textit{Astronomy \& Astrophysics}, 681, L5.

\bibitem[Barbieri \& D\'emoulin(2025)]{Barbieri2025}
Barbieri, L. \& D\'emoulin, P. (2025).
\newblock Kinetic collisionless model of the solar transition region and corona with spatially intermittent heating.
\newblock \textit{Astronomy \& Astrophysics}, 704, A84.

\bibitem[Boldyrev \& Horaites(2019)]{Boldyrev2019}
Boldyrev, S. \& Horaites, K. (2019).
\newblock Kinetic theory of the electron strahl in the solar wind.
\newblock \textit{Monthly Notices of the Royal Astronomical Society}, 489, 3412.

\bibitem[Chiuderi \& Chiuderi Drago(2004)]{ChiuderiDrago2004}
Chiuderi, C. \& Chiuderi Drago, F. (2004).
\newblock Effect of suprathermal particles on the quiet Sun radio emission.
\newblock \textit{Astronomy \& Astrophysics}, 422, 331.

\bibitem[Cranmer(2014)]{Cranmer2014}
Cranmer, S.~R. (2014).
\newblock Suprathermal electrons in the solar corona: can nonlocal transport explain Heliospheric charge states?
\newblock \textit{The Astrophysical Journal Letters}, 791, L31.

\bibitem[Del~Zanna \& Mason(2018)]{DelZanna2018}
Del~Zanna, G. \& Mason, H.~E. (2018).
\newblock Solar {UV} and {X}-ray spectral diagnostics.
\newblock \textit{Living Reviews in Solar Physics}, 15, 5.

\bibitem[Dorelli \& Scudder(2003)]{Dorelli2003}
Dorelli, J.~C. \& Scudder, J.~D. (2003).
\newblock Electron heat flow in the solar corona: implications of non-Maxwellian velocity distributions.
\newblock \textit{Journal of Geophysical Research}, 108, 1294.

\bibitem[Dud\'{\i}k et~al.(2012)]{Dudik2012}
Dud\'{\i}k, J., Ka\v{s}parov\'a, J., Dzi\v{f}\v{c}\'akov\'a, E., Karlick\'y, M., \& Mackovjak, \v{S}. (2012).
\newblock The non-Maxwellian continuum in the X-ray, UV, and radio range.
\newblock \textit{Astronomy \& Astrophysics}, 539, A107.

\bibitem[Dud\'{\i}k et~al.(2014)]{Dudik2014}
Dud\'{\i}k, J., Del~Zanna, G., Mason, H.~E., \& Dzi\v{f}\v{c}\'akov\'a, E. (2014).
\newblock Signatures of the non-Maxwellian $\kappa$-distributions in optically thin line spectra. I. Theory and synthetic Fe~IX--XIII spectra.
\newblock \textit{Astronomy \& Astrophysics}, 570, A124.

\bibitem[Dud\'{\i}k et~al.(2015)]{Dudik2015}
Dud\'{\i}k, J., Mackovjak, \v{S}., Dzi\v{f}\v{c}\'akov\'a, E., et~al. (2015).
\newblock Spectroscopic diagnostics of warm coronal loops observed by SDO/AIA: evidence for $\kappa$-distributions.
\newblock \textit{The Astrophysical Journal}, 807, 123.

\bibitem[Dud\'{\i}k et~al.(2017)]{Dudik2017}
Dud\'{\i}k, J., Dzi\v{f}\v{c}\'akov\'a, E., Meyer-Vernet, N., et~al. (2017).
\newblock Nonequilibrium processes in the solar corona, transition region, flares, and solar wind (invited review).
\newblock \textit{Solar Physics}, 292, 100.

\bibitem[Dulk(1985)]{Dulk1985}
Dulk, G.~A. (1985).
\newblock Radio emission from the Sun and stars.
\newblock \textit{Annual Review of Astronomy and Astrophysics}, 23, 169.

\bibitem[Dzi\v{f}\v{c}\'akov\'a et~al.(2018)]{Dzifcakova2018}
Dzi\v{f}\v{c}\'akov\'a, E., Zemanov\'a, A., Dud\'{\i}k, J., \& Mackovjak, \v{S}. (2018).
\newblock Spectroscopic diagnostics of the non-Maxwellian $\kappa$-distributions using {SDO/EVE} observations of the 2012 {M}arch 7 {X}-class flare.
\newblock \textit{The Astrophysical Journal}, 853, 158.

\bibitem[Gautam et~al.(2024)]{Gautam2024}
Gautam, S.~P., Adhikari, L., Zank, G.~P., Silwal, A., \& Zhao, L. (2024).
\newblock Solar Cycle Dependence of the Turbulence Cascade Rate at 1 au.
\newblock \textit{The Astrophysical Journal}, 968, 12.

\bibitem[Gibson et~al.(2016)]{Gibson2016}
Gibson, S.~E., Kucera, T.~A., White, S.~M., et~al. (2016).
\newblock FORWARD: A Toolset for Multiwavelength Coronal Magnetometry.
\newblock \textit{Frontiers in Astronomy and Space Sciences}, 3, 8.

\bibitem[Fleishman \& Kuznetsov(2014)]{Fleishman2014}
Fleishman, G.~D. \& Kuznetsov, A.~A. (2014).
\newblock Theory of gyroresonance and free-free emissions from non-Maxwellian quasi-steady-state electron distributions.
\newblock \textit{The Astrophysical Journal}, 781, 77.

\bibitem[Halekas et~al.(2020)]{Halekas2020}
Halekas, J.~S., Whittlesey, P., Larson, D.~E., et~al. (2020).
\newblock Electrons in the young solar wind: first results from Parker Solar Probe.
\newblock \textit{The Astrophysical Journal Supplement Series}, 246, 22.

\bibitem[Landi(2007)]{Landi2007}
Landi, E. (2007).
\newblock Ion temperatures in the quiet solar corona.
\newblock \textit{The Astrophysical Journal}, 663, 1363.

\bibitem[Landi \& Pantellini(2001)]{Landi2001}
Landi, S. \& Pantellini, F.~G.~E. (2001).
\newblock On the temperature profile and heat flux in the solar corona: kinetic simulations.
\newblock \textit{Astronomy \& Astrophysics}, 372, 686.

\bibitem[Livadiotis \& McComas(2009)]{Livadiotis2009}
Livadiotis, G. \& McComas, D.~J. (2009).
\newblock Beyond kappa distributions: exploiting Tsallis statistical mechanics in space plasmas.
\newblock \textit{Journal of Geophysical Research}, 114, A11105.

\bibitem[Lorin\v{c}\'{\i}k et~al.(2020)]{Lorincik2020}
Lorin\v{c}\'{\i}k, J., Dud\'{\i}k, J., Del~Zanna, G., Dzi\v{f}\v{c}\'akov\'a, E., \& Mason, H.~E. (2020).
\newblock Plasma Diagnostics from Active Region and Quiet-Sun Spectra Observed by Hinode/EIS: Quantifying the Departures from a Maxwellian Distribution.
\newblock \textit{The Astrophysical Journal}, 893, 34.

\bibitem[Mercier \& Chambe(2015)]{Mercier2015}
Mercier, C. \& Chambe, G. (2015).
\newblock Electron density and temperature in the solar corona from multifrequency radio imaging.
\newblock \textit{Astronomy \& Astrophysics}, 583, A101.

\bibitem[Meyer-Vernet et~al.(1995)]{MeyerVernet1995}
Meyer-Vernet, N., Moncuquet, M., \& Hoang, S. (1995).
\newblock Temperature inversion in the Io plasma torus.
\newblock \textit{Icarus}, 116, 202.

\bibitem[Owocki \& Scudder(1983)]{Owocki1983}
Owocki, S.~P. \& Scudder, J.~D. (1983).
\newblock The effect of a non-Maxwellian electron distribution on oxygen and iron ionization balances in the solar corona.
\newblock \textit{The Astrophysical Journal}, 270, 758.

\bibitem[Pierrard \& Lazar(2010)]{Pierrard2010}
Pierrard, V. \& Lazar, M. (2010).
\newblock Kappa distributions: theory and applications in space plasmas.
\newblock \textit{Solar Physics}, 267, 153.

\bibitem[Pierrard \& Lemaire(1996)]{Pierrard1996}
Pierrard, V. \& Lemaire, J. (1996).
\newblock Lorentzian ion exosphere model.
\newblock \textit{Journal of Geophysical Research}, 101, 7923.

\bibitem[Scudder(1992{\it a})]{Scudder1992a}
Scudder, J.~D. (1992{\it a}).
\newblock On the causes of temperature change in inhomogeneous low-density astrophysical plasmas.
\newblock \textit{The Astrophysical Journal}, 398, 299.

\bibitem[Scudder(1992{\it b})]{Scudder1992b}
Scudder, J.~D. (1992{\it b}).
\newblock Why all stars should possess circumstellar temperature inversions.
\newblock \textit{The Astrophysical Journal}, 398, 319.

\bibitem[Scudder(2019)]{Scudder2019}
Scudder, J.~D. (2019).
\newblock Steady Electron Runaway Model SERM: Astrophysical Alternative for the Maxwellian Assumption.
\newblock \textit{The Astrophysical Journal}, 885, 138.

\bibitem[Sharma \& Oberoi(2020)]{Sharma2020}
Sharma, R. \& Oberoi, D. (2020).
\newblock Propagation Effects in Quiet Sun Observations at Meter Wavelengths.
\newblock \textit{The Astrophysical Journal}, 903, 126.

\bibitem[Shoub(1983)]{Shoub1983}
Shoub, E.~C. (1983).
\newblock Invalidity of local thermodynamic equilibrium for electrons in the solar transition region. I.
\newblock \textit{The Astrophysical Journal}, 266, 339.

\bibitem[\v{S}tver\'ak et~al.(2009)]{Stverak2009}
\v{S}tver\'ak, \v{S}., Maksimovic, M., Tr\'avn\'{\i}\v{c}ek, P.~M., et~al. (2009).
\newblock Radial evolution of nonthermal electron populations in the low-latitude solar wind.
\newblock \textit{Journal of Geophysical Research}, 114, A05104.

\bibitem[Thejappa \& MacDowall(2008)]{Thejappa2008}
Thejappa, G. \& MacDowall, R.~J. (2008).
\newblock Effects of scattering on radio emission from the quiet Sun at low frequencies.
\newblock \textit{The Astrophysical Journal}, 676, 1338.

\bibitem[Vocks et~al.(2008)]{Vocks2008}
Vocks, C., Mann, G., \& Rausche, G. (2008).
\newblock Formation of suprathermal electron distributions in the quiet solar corona.
\newblock \textit{Astronomy \& Astrophysics}, 480, 527.

\bibitem[Vocks et~al.(2018)]{Vocks2018}
Vocks, C., Mann, G., Breitling, F., et~al. (2018).
\newblock LOFAR observations of the quiet solar corona.
\newblock \textit{Astronomy \& Astrophysics}, 614, A54.

\end{thebibliography}

\end{document}